\documentclass[aps,prl,reprint,superscriptaddress,showpacs,preprintnumbers,amsmath,amssymb]{revtex4-2}
\usepackage[dvipdfmx]{graphicx}
\usepackage{dcolumn}   
\usepackage{bm}        
\usepackage{amssymb, amsmath, amsfonts}   
\usepackage{lipsum}        
\usepackage{color}
\usepackage{multirow}
\DeclareGraphicsExtensions{{.pdf}}
\begin{document}
\newcommand{\noter}[1]{{\color{red}{#1}}}
\newcommand{\noteb}[1]{{\color{blue}{#1}}}
\newcommand{\field}{\left( \boldsymbol{r}\right)}
\newcommand{\paren}[1]{\left({#1}\right)}
\newcommand{\vect}[1]{\boldsymbol{#1}}
\newcommand{\uvect}[1]{\tilde{\boldsymbol{#1}}}
\newcommand{\vdot}[1]{\dot{\boldsymbol{#1}}}
\newcommand{\vder}{\boldsymbol{\nabla}}
\widetext

\title{Scale Separation of Shear-induced Criticality in Glasses
}

\author{Norihiro Oyama}
\email{Norihiro.Oyama.vb@mosk.tytlabs.co.jp}
\affiliation{Toyota Central R\&D Labs., Inc., {Nagakute 480-1192}, Japan}

\author{Takeshi Kawasaki}
\affiliation{Department of Physics, Nagoya University, Nagoya 464-8602, Japan}

\author{Kang Kim}
\affiliation{Division of Chemical Engineering, Graduate School of Engineering Science, Osaka University, Osaka 560-8531, Japan}

\author{Hideyuki Mizuno}
\affiliation{Graduate School of Arts and Sciences, The University of Tokyo, Tokyo 153-8902, Japan}
\date{\today}
\begin{abstract}
In a sheared steady state, glasses reach a nonequilibrium criticality called yielding.
In this letter, we report that the qualitative nature of this nonequilibrium critical phenomenon depends on the details of the system and that responses and fluctuations are governed by different critical correlation lengths in specific situations.
This scale separation of critical lengths arises when the screening of elastic propagation of mechanical signals is not negligible.
We also explain that the impact of the screening effects is crucially determined by the microscopic dissipation mechanism.
\end{abstract}
\maketitle
%
\emph{Introduction.}---
In athermal (zero-temperature) sheared glasses, fluidization proceeds through many local plastic events~\cite{ML2004a,ML2004b,ML2006,LC2007PRE,LC2009,FSE_stress2,Karmakar2010,Martens2011,Roy,Ozawa2017,Zhang2017,Karimi,Oyama2019PRL,Saitoh2019,Ferrero2019,Oyama2021PRE,Oyama,Oyama2021G}.
In a steady state, in particular, the elementary processes of these plastic events, so-called shear transformations (STs), tend to form avalanches~\cite{ML2004b,ML2006,Oyama}.
Such avalanches can sometimes span the whole system and result in one class of nonequilibrium criticality~\cite{Sethna2001} called yielding~\cite{Lin2014,Ferrero2019,Oyama}.
The criticality is typically reflected by {stress response}:
the average stress $\langle\sigma\rangle$ obeys a critical phenomenon-like functional form called the Herschel-Bulkley (HB) law $\langle\sigma\rangle-\sigma_{\rm Y}\sim\dot{\gamma}^n$~\cite{HB1926} and exhibits a finite size effect
~\cite{LC2009,FSE_stress2,Roy,Oyama}.
Here, $\sigma_{\rm Y}$ is the yield stress, $\dot{\gamma}$ is the applied strain rate, and $n$ is the HB exponent.
This yielding criticality is also characterized by scaling ansatzes~\cite{Lin2014,Oyama}
$\xi\sim|\langle\sigma\rangle-\sigma_{\rm Y}|^{-\nu}$ and $\dot{\gamma}\sim|\langle\sigma\rangle-\sigma_{\rm Y}|^{\beta}$, where $\xi$ is the critical correlation length of avalanches and $\nu$ and $\beta$ are critical exponents.
From {the statistical tilt symmetry} of the governing equation, we can show that the exponent $\nu$ has a hyperscaling relation $\nu=1/(d-d_f)$ with the fractal dimension $d_f$ of the geometric structure of avalanches~\cite{Lin2014}.
The fractal dimension $d_f$ and the yield stress $\sigma_{\rm Y}$\footnote{$\sigma_{\rm Y}$ is the value in the thermodynamic limit and can be estimated from the system-size dependence of $\protect\langle\sigma\protect\rangle$ under quasistatic shear~\cite{Lin2014,Oyama}.} can be determined by simulations under quasistatic shear~\cite{Lin2014,Oyama2021PRE,Oyama}.
From the above scaling relation, we can further determine $\nu$.

Although there was no way to measure the remaining important parameter $\beta$ systematically in particulate systems, we recently found that we can obtain $\beta$ from the average number of STs occurring simultaneously, $\langle N_{\rm ST}\rangle$.
We can describe $\langle N_{\rm ST}\rangle$ as:
\begin{align}
    \langle N_{\rm ST}\rangle\sim N_{\rm ava}\cdot N_{\rm ST/ava}\sim N\cdot \dot{\gamma}^{1/\beta},\label{eq:N_dag}
\end{align}
where, $N_{\rm ava}$ and $N_{\rm ST/ava}$ represent the number of avalanches in the system and the number of STs in each avalanche respectively.
By definition, they can be expressed as $N_{\rm ava}\sim (L/\xi)^d$ and $N_{\rm ST/ava}\sim  \xi^{d_f}$~\cite{Oyama} (see {Supplementary Material (SM)} for a detailed explanation).
Using Eq.~\ref{eq:N_dag}, we can determine $\beta$ from the $\dot{\gamma}$ dependence of the average number density of STs $\langle n_{\rm ST}\rangle\equiv\langle N_{\rm ST}\rangle/N$.
Furthermore, we found that instantaneous normal modes with imaginary frequencies (we call them Im-INMs) correspond to activated STs that are causing plastic deformations~\cite{Oyama} and thus $N_{\rm ST}$ can be estimated from the number of Im-INMs.
Instantaneous normal modes are obtained as the eigenmodes of the Hessian matrix of the total potential energy of an instantaneous configuration~\cite{INM1,INM2,Gezelter1997,INM4,INM5}, which is available in particulate systems.
{Therefore, the important parameters to characterize yielding criticality, $\sigma_Y$, $\nu$, and $\beta$, can all be determined systematically.}
{In ref.~\cite{Oyama}, we demonstrated that the estimated parameters describe well the criticality of the numerically observed stress and established the validity of the concept of yielding criticality.}

Moreover, in sheared glasses, the dynamics of constituent particles become diffusive even under athermal conditions~\cite{LC2007PRE,LC2009,Roy,Martens2011,Karimi}.
If we quantify diffusive motions by the strain-based diffusion constant $\hat{D}$, they also exhibit criticality as $\hat{D}\sim \dot{\gamma}^{-n_{\hat{D}}}$. 
Here, we introduced a critical exponent $n_{\hat{D}}$ and defined the diffusion constant as $\hat{D}\equiv \hat{\Delta}^2(\gamma_t\to \infty)$ using the strain-based mean-squared displacements in the $y$-direction, $\hat{\Delta}^2(\gamma_t)\equiv \langle\frac{1}{N}\sum_i(y_i(0)-y_i(t))^2\rangle/\gamma_t$.
We assume that the system is two-dimensional (2D)~\cite{LC2007PRE,LC2009,Roy,Martens2011} and that the shear is applied in the $x$-direction.
$\gamma_t\equiv \dot{\gamma}t$ is the strain applied during a time interval $t$.
The precise measurements of the exponent $n_{\hat{D}}$, including the confirmation by finite size scaling (FSS), were performed in refs.~\cite{LC2009,Roy}.
{In particular, in ref.~\cite{LC2009}, based on a general phenomenological discussion, a theoretical prediction for the exponent was given as $n_{\hat{D}}=1/2$, which is in line with the numerical result.}
This prediction was considered not to depend on the system details according to the general nature of the theoretical treatment.
However, ref.~\cite{Roy} reported a largely different value of $n_{\hat{D}}=1/3$ under a different numerical setup.
We lack an understanding of the cause of this unexpected diversity of $n_{\hat{D}}$.

In this Letter, we study the cause of the diversity in $n_{\hat{D}}$ by means of molecular dynamics simulations of 2D sheared glasses and the scaling argument based on the recently established yielding criticality.
{Our main finding is that there are not only quantitative differences in the values of exponents but also qualitative differences in the property of the criticality itself.}
We first found that $n_{\hat{D}}$ is inversely proportional to $\beta$. 
Since the diversity of $\beta$ is already known~\cite{shear_review1,Shear_review2}, this inverse proportionality explains the diversity in $n_{\hat{D}}$ reported thus far~\cite{LC2009,Roy}.
{By comparing the results for two systems with qualitatively different microscopic dissipation mechanisms, we also show that the diverging correlation length governing the criticality of diffusion differs from that governing stress in certain situations.
Whether such a scale separation is present or not causes a major difference in $n_{\hat{D}}$ even when $\beta$ is nearly the same.
We emphasize that critical phenomena involving two length scales are {very rare} and have been reported only in a few examples, such as jamming~\cite{two_lengths} and quantum phase transitions~\cite{Shao2016}.}
We further clarified the physical meaning of the second critical correlation length and the reason why such scale separation could be observed only under certain conditions.

\emph{Numerical setups.}---
We conduct molecular dynamics simulations of 2D ($d=2$) glasses under external shear.
The interparticle interaction follows from the Lennard-Jones potential with smoothing terms~\cite{Oyama,Oyama2021PRE} which ensure that the potential and force smoothly tend to zero at the cutoff distance $r^c_{ij}=1.3d_{ij}$, where $d_{ij}$ determines
the interaction range between particles $i$ and $j$.
{To avoid crystallization, we consider a 50:50 mixture of two types of particles with different sizes but with the same mass $m=1.0$.
The interaction ranges for different pairs of particle types are $d_{\rm SS}=5/6$, $d_{\rm SL}=1.0$ and $d_{\rm LL}=7/6$, where subscripts $S$ and $L$ distinguish particle types.
The energy scale $\epsilon_{ij}=\epsilon = 1.0$ is constant for all particle pairs.
The physical variables reported in this letter are all nondimensionalized by $d_{SL}$, $m$ and $\epsilon$.
The number density $\rho$ is set to be $\rho=N/L^2\sim 1.09$, where $N$ is the number of particles and $L$ is the corresponding linear dimension of the system.

{We applied the shear at different rates in the range of $2\times 10^{-5}\le\dot{\gamma}\le 2\times 10^{-2}$.
{At every simulation step, we first impose affine simple shear of strain $\Delta \gamma$ in the $x$-direction and then calculate the nonaffine dynamics for a time interval of $\Delta t$ by integrating the
equations of motion under the Lees-Edwards boundary conditions~\cite{Allen1987}.}
{The shear rate is expressed as $\dot{\gamma}=\Delta\gamma/\Delta t$.
To make the strain resolution constant, we fix $\Delta
\gamma$ to be $1.0\times 10^{-7}$ and control $\dot{\gamma}$ by changing the time step $\Delta t$.}

We considered two types of systems, Systems A and B, that have different microscopic dissipation mechanisms:
\begin{itemize}
    \item System A: ``Contact'' damping\\
    In System A, dissipative interparticle forces are introduced as $\boldsymbol{f}^{\rm visc.}_i=\frac{m}{\tau}\sum_j\phi(r_{ij})(\delta\boldsymbol{v}_j-\delta\boldsymbol{v}_i)$~\cite{LC2009}.
    Here, $\phi(r_{ij})\propto 1-2(r_{ij}/r_{ij}^{\rm C})^4+(r_{ij}/r_{ij}^{\rm C})^8$ is a smoothing function, $\delta\boldsymbol{v}_i$ is the nonaffine velocity, and $\tau=0.2$ is the dissipation time scale.
    This type of dissipation selectively damps high-wavenumber local relative motions~\cite{Maloney2008}.
    \item System B: Stokes drag\\
    In System B, the dissipation is modeled through the Stokes drag force as $\boldsymbol{f}^{\rm
drag}_{
  i}=-\Gamma \delta\boldsymbol{v}_i$~\cite{Lemaitre2006,Salerno2012}, where the damping coefficient $\Gamma$ is set to unity.
  \end{itemize}}

Averages are denoted by angular brackets $\langle\cdot\rangle$ and they are calculated over steady-state data ($1\le\gamma\le 20$) and $8$ independent samples.
Initial configurations are all generated by the minimization of the potential energy of totally random structures.
We ignore the thermal fluctuations (athermal situation).}


\begin{figure}
\includegraphics[width=\linewidth]{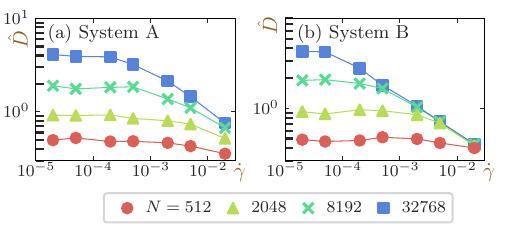}
\caption{
{Strain-based diffusion constant $\hat{D}$ as a function of $\dot{\gamma}$. 
Results for (a) System A and (b) System B.
}\label{fig:diffusion}
}
\end{figure}

\emph{Scaling argument for diffusion constant.}---
We plot the strain-based diffusion constant $\hat{D}$ for different system sizes $N$ as functions of the shear rate $\dot{\gamma}$ in Fig.~\ref{fig:diffusion}(a,b).
In the low $\dot{\gamma}$ regime below a system-size dependent threshold $\dot{\gamma}_{C_{\hat{D}}}$, $\hat{D}$ plateaus at $\hat{D}_0$.
The plateau value $\hat{D}_0$ also linearly depends on the system linear dimension $L$ as $\hat{D}_0\sim L$~\footnote{Consistently with the observations, a scaling relation $\hat{D}_0\sim L^{d-d_f}$ was proposed~\cite{Karimi}.}.
In the high-rate limit, on the other hand, $\hat{D}$ seems to obey characteristic power laws.
All these behaviors are common to both Systems A and B and consistent with reports in refs~\cite{LC2009,Roy}.

To describe the critical behavior of $\hat{D}$ in the framework of the yielding criticality, we introduce another scaling ansatz with a new exponent $\alpha$ as $\hat{D}\sim\Delta\sigma^{-\alpha}\sim\dot{\gamma}^{-\alpha/\beta}$. 
Introducing a scaling function $f_{\hat{D}}(x)$, we obtain
$\hat{D}\cdot\dot{\gamma}_{C_{\hat D}}^{\alpha/\beta}=(\dot{\gamma}/\dot{\gamma}_{C_{\hat D}})^{-\alpha/\beta}f_{\hat{D}}(\dot{\gamma}/\dot{\gamma}_{C_{\hat D}})$.
On the other hand, we can also describe $\hat{D}$ using the critical correlation length ${\xi}$ as $\hat{D}\sim {\xi}^{\alpha/\nu}$.
Because $\hat{D}$ becomes $\hat{D}_0$ at $\dot{\gamma}=\dot{\gamma}_{C_{\hat{D}}}$, we obtain a relation $L^{\alpha/\nu}\sim\hat{D}_0\sim L$ and $\dot{\gamma}_{C_{\hat D}}\sim L^{-\beta/{\nu}}$ ({see SM} for details).
From all these relations, we obtain $\alpha=\nu$ and 
\begin{align}
\hat{D}/L=(L\cdot\dot{\gamma}^{\nu/\beta})^{-1}f_{\hat{D}}(\dot{\gamma}/\dot{\gamma}_{C_{\hat{D}}}).\label{eq:D_scaling}
\end{align}
Therefore with correct values of $\nu$ and $\beta$, we expect the FSS according to this equation to collapse curves of $\hat{D}$ for different system sizes. 
Additionally, by comparing the exponents, we obtain a relation $n_{\hat{D}}=\nu/\beta$.
These results mean that the criticality of $\hat{D}$ can be described only by $\beta$ and $\nu$, the exponents introduced to describe the criticality of the stress response.

\tabcolsep = 7pt
\renewcommand{\arraystretch}{1.4}
\begin{table*}[bth]
  {\caption{\label{table:scalings}Summary of exponents and scaling ansatzes}}
  \centering
  \begin{tabular}{cccccc}
    \hline
     {Exponent} & {Corresponding variable} & {Definition} &{Measurement} &System A &System B\\
    \hline 
    \hline 
    $\beta$ & Shear rate &$\dot{\gamma}\sim\Delta\sigma^\beta$ &INMs analysis~\cite{Oyama}&1.39&1.32\\
    \hline 
    $d_f$ &Avalanche size &$S\sim\xi^{d_f}$ &Quasistatic simulations~\cite{Oyama2021PRE}&\multicolumn{2}{c}{1.03}\\
    \hline 
    $\nu$ & Avalanche correlation length &$\xi\sim\Delta\sigma^{-\nu}$ &$\nu=1/(d-d_f)$\cite{Lin2014}&\multicolumn{2}{c}{1.04}\\
    \hline 
    $\tilde{\nu}$ & ST correlation length &$\zeta\sim\Delta\sigma^{-\tilde{\nu}}$ &$\tilde{\nu}=1/d$&\multicolumn{2}{c}{0.50}\\
    \hline 
    {$n_{\hat{D}}$} &{Diffusion constant} &{$\hat{D}\sim\dot{\gamma}^{-n_{\hat{D}}}$} &$n_{\hat{D},A}=\nu/\beta_A$, $n_{\hat{D},B}=\tilde{\nu}/\beta_B$  &{0.74} &{0.38}\\
    \hline
  \end{tabular}
\end{table*}

\begin{figure}
\includegraphics[width=\linewidth]{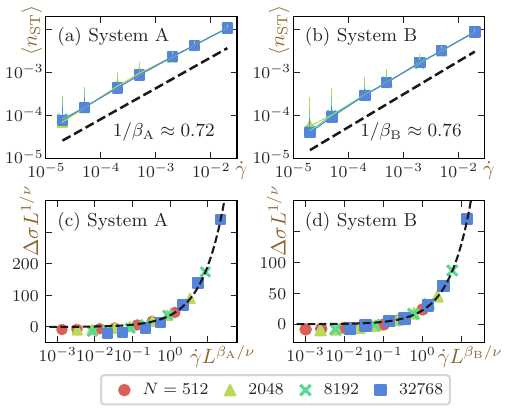}
\caption{
(a,b) Average number density of STs $\langle n_{\rm ST}\rangle$ as a function of the shear rate $\dot{\gamma}$. 
{The dashed lines are the fitting results using all data points and have slopes of $1/\beta_{\rm A}\approx{0.72}$ and $1/\beta_{\rm B}\approx{0.76}$.}
(c,d) $\Delta\sigma\equiv\langle\sigma\rangle - \sigma_{\rm Y}$ as a function of shear rate with finite size scaling.
{The dashed lines show the HB law: $\Delta\sigma\sim\dot{\gamma}^{1/\beta_{\rm A/B}}$.}
Left: Results for System A. Right: Results for System B.
\label{fig:stress_response}
}
\end{figure}

\emph{Measurement of exponents and yield stress.---}
As explained above, we can extract the precise values of $\beta$ from $\langle n_{\rm ST}\rangle$, the average number density of STs.
In Fig.~\ref{fig:stress_response}(a,b), we plot $\langle n_{\rm ST}\rangle${ estimated by the number of Im-INMs~\cite{Oyama}} as a function of $\dot{\gamma}$ for Systems A and B.
As expected from Eq.~\ref{eq:N_dag}, $\langle n_{\rm ST}\rangle$ does not show any system-size dependence, and the results for all system sizes obey a master power-law curve for both systems.
From the slopes, we can determine the values of $\beta$ as {$\beta_{\rm A}\approx 1.39$} and $\beta_{\rm B}\approx 1.32$, where subscripts ${\rm A/B}$ distinguish the system of interest (we summarize all values of critical exponents in Table~\ref{table:scalings}.). 
We note that although $\beta_{\rm A}$ and $\beta_{\rm B}$ happened to be close, the value of the HB exponent $n=1/\beta$ generally depends on the details of the system, with values as varied as $0.2\le n\le 0.8$ being reported~\cite{shear_review1,Shear_review2}.

Moreover, we can determine $\sigma_{\rm Y}$ and $d_f$ from quasistatic simulations~\cite{Lin2014}. For our systems, we measured $\sigma_{\rm Y}\approx 3.75$ and $d_f\approx 1.03$ and through the relation $\nu=1/(d-d_f)$, we obtain $\nu\approx 1.04$~\cite{Oyama,Oyama2021PRE}.
Since these values are solely determined by quasistatic dynamics, they are shared by both Systems A and B.
We demonstrate the success of FSS of $\Delta\sigma\equiv \langle\sigma\rangle - \sigma_{\rm Y}$ with these parameters in Fig.~\ref{fig:stress_response}(c,d) for Systems A and B respectively (see SM for details).
These results guarantee the correctness of exponents $\beta_{\rm A/B}$ and $\nu$.

\begin{figure}
\includegraphics[width=\linewidth]{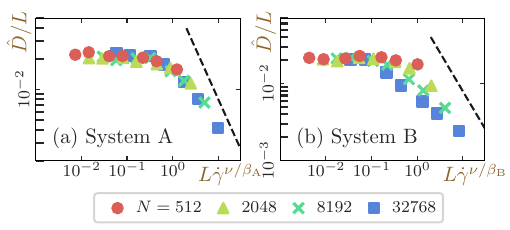}
\caption{
{Finite size scaling of strain-based diffusion constant $\hat{D}$ using exponent $\nu\approx 1.04$ associated with the avalanche correlation length $\xi$.
Dashed lines represent the slope of $-1$ expected from the scaling argument.
Results for (a) System A and (b) System B.
}\label{fig:xi_scaling}
}
\end{figure}

{We now try an FSS of $\hat{D}$ using the obtained exponents and Eq.~\ref{eq:D_scaling}: the results are shown in Fig.~\ref{fig:xi_scaling}.}
As shown in Fig.~\ref{fig:xi_scaling}(a), for System A, the results for different $N$ collapse.
This indicates that $\hat{D}$ is governed by $\xi$, and thus $n_{\hat{D},{\rm A}}$ is estimated as $n_{\hat{D},{\rm A}}=\nu/\beta_A\approx 0.72$.
{We note that, in systems of both refs.~\cite{LC2009,Roy}, the criticalities of $\hat{D}$ were described by $\xi$ as well, as explained in detail {in SM}.
This means that the variation in $n_{\hat{D}}$ among these systems simply derives from that in $n=1/\beta$ ($n=1/2$~\cite{LC2009} and 1/3~\cite{Roy}).}

On the other hand, as shown in Fig.~\ref{fig:xi_scaling}(b), the FSS is not successful for System B: the data in the high-shear-rate scaling regime vary significantly among different $N$.
{The failure of this attempt is due to the inadequacy of the implicit assumption in Eq.~2 that the criticality of $\hat{D}$ and $\langle\sigma\rangle$ are governed by the same correlation length $\xi$.
Below, we explain that there exists another critical correlation length and that the second length governs the criticality of $\hat{D}$ in System B.}


\emph{Existence of another length scale.}---
In the phenomenological discussion in refs.~\cite{LC2009,Karimi},
$\hat{D}$ was described by a superposition of Eshelby fields~\cite{Picard2004} induced by all STs in the system ({see SM} for a brief summary of the theoretical background). 
Importantly, to reproduce the critical finite size effect, the correlation between STs over the length scale $\xi$ was crucial.
In this phenomenological consideration, the effect of each Eshelby field was assumed to propagate throughout the whole system via an elastic field.

When multiple STs are excited simultaneously, however, this assumption may not always hold.
If we consider the displacement of a particle induced by Eshelby fields, the local plastic motion of nearby STs can screen the elastic propagation of Eshelby fields emitted by distant STs.~\cite{screen}.
This effectively divides the whole system into \emph{purely elastic} subsystems.
Therefore, when such screening effects cannot be ignored, the linear dimension of such elastic subdomains plays a major role in determining the diffusivity.
{We name such a length scale $\zeta$ and assume another scaling ansatz for it as $\zeta\sim\Delta\sigma^{-\tilde{\nu}}$, introducing another exponent $\tilde{\nu}$.
{If the criticality of $\hat{D}$ is governed by $\zeta$,} we obtain $\alpha=\tilde{\nu}$ and $\hat{D}/L=(L\cdot\dot{\gamma}^{\tilde{\nu}/\beta})^{-1}g_{\hat{D}}(\dot{\gamma}/\dot{\gamma}_{C_{\hat{D}}})$ from the same derivation of Eq.~\ref{eq:D_scaling} ({see also SM}).}

\begin{figure}
\includegraphics[width=\linewidth]{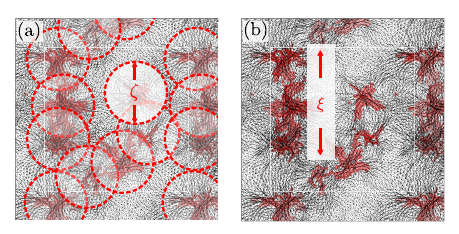}
\caption{
Schematic picture of (a) ST correlation length $\zeta$ and (b) avalanche correlation length $\xi$.
Mobile particles ({see SM} for precise definition) of all Im-INMs that correspond to active STs are highlighted in red.
The results from a system with $N=2048$ under shear of the rate $\dot{\gamma}=2\times 10^{-3}$ are shown.
The copied images due to the periodic boundary conditions are also visualized around the original computational domain with lighter colors.
Black arrows depict the eigenvectors.
  \label{fig:xi_zeta}}
\end{figure}

To consider the finite size scaling of $\hat{D}$ by $\zeta$, we need to estimate $\tilde{\nu}$. For this, $\langle N_{\rm ST}\rangle$ can be utilized again.
{The average volume of elastic subdomains corresponds to the average volume occupied by a single ST (see a schematic picture in Fig.~\ref{fig:xi_zeta}(a): we call $\zeta$ the ST correlation length hereafter).}
Thus, the exponent $\tilde{\nu}$ can be determined by considering a situation where $\langle N_{\rm ST}\rangle\sim\zeta^d\dot{\gamma}^{1/\beta}\sim 1$ (from Eq.~\ref{eq:N_dag} with $L=\zeta$).
We obtain $\tilde{\nu}=1/d$ from this relation ({see SM} for details).
Since the two exponents, $\nu\approx 1.0$ and $\tilde{\nu}=1/d=1/2$, are largely different, the corresponding lengths $\xi$ and $\zeta$ are different in nature in view of the critical phenomena.
{As a reference, we show the sketch for the avalanche correlation length $\xi$ in Fig.~\ref{fig:xi_zeta}(b).
As shown here, $\xi$ corresponds to, by definition, the overall spanning length of avalanches formed by STs ({see SM} for a precise definition).}

\begin{figure}
\includegraphics[width=\linewidth]{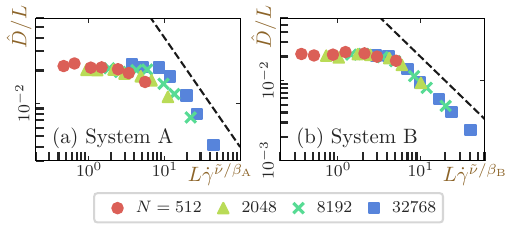}
\caption{
Finite size scaling of the strain-based diffusion constant $\hat{D}$  using the exponent $\tilde{\nu}=1/d$ associated with the ST correlation length $\zeta$.
Dashed lines represent the slope of $-1$ expected from the scaling argument.
(a) Results for System A. (b) Results for System B.
  \label{fig:zeta_scaling}}
\end{figure}

In Fig.~\ref{fig:zeta_scaling}(a), we plot the results of the FSS of $\hat{D}$ using the critical exponent $\tilde{\nu}$ for System A: the scaling is obviously not successful.
This failure allows us to reconfirm that the criticality of $\hat{D}$ is governed solely by $\xi$, not by $\zeta$.
In Fig.~\ref{fig:zeta_scaling}(b), we tried the same FSS for System B.
In this case, we see a perfect collapse of results for different $N$ and can conclude that $\hat{D}$ is governed by $\zeta$.
In other words, in this system, separation of the critical correlation lengths is observed between criticalities of $\langle\sigma\rangle$ (a measure of response) and $\hat{D}$ (a measure of fluctuations):
{Since fluctuations are locally determined, they are affected by screening effects, whereas the response is globally determined by the total spanning length of avalanches and is therefore independent of screening effects.}
{Because this scale separation is present only in System B, even though $\beta_{\rm A}\approx 1.39$ and $\beta_{\rm B}\approx 1.32$ are close, $n_{\hat{D},{\rm A}}\approx 0.72$ and $n_{\hat{D},{\rm B}}\approx 0.38$ are largely different.
}
{We emphasize that the presence of multiple critical correlation lengths is a rare property, observed only for limited phenomena such as jamming~\cite{two_lengths} and quantum phase transitions~\cite{Shao2016}.}

An important question remains: Why is the screening effect negligible in System A?
This is likely because high-wavenumber local relative motions are overdamped in this system~\cite{Maloney2008}.
Because of this feature, local motions resulting from the excitation of STs are suppressed and the screening effect of elastic wave propagation becomes very weak.
We note that, as we explain in {detail in SM}, reinterpretation of reported values of exponents in ref.~\cite{LC2009,Roy} indicates that the scale separation is also negligible in the systems in these studies.
The high-wavenumber motions are overdamped in those systems in refs.~\cite{LC2009,Roy} as well, which is consistent with the discussion in this paragraph.

\emph{Summary and overview.}---
In this letter, by means of molecular dynamics simulations of sheared 2D glasses, we studied the origin of the diversity of exponent $n_{\hat{D}}$ characterizing the criticality of the diffusion coefficient, which had remained previously unclear.
{We found that the diversity in $n_{\hat{D}}$ was caused not only by quantitative but also by qualitative differences: whether the scale separation of the critical correlation lengths of response and fluctuation is present or not.}
We also revealed that the screening effect of elastic waves, which arises when microscopic dissipation does not completely damp high-wavenumber local dynamics, is responsible for the emergence of such scale separation.

{
The authors thank Masanari Shimada for fruitful discussions.
This work was financially supported by the JST FOREST Program (Grant No. JPMJFR212T) and JSPS KAKENHI {Grant Numbers 20H01868, 22H04472, 22K03543, 23H04495, 23H04503, and JP22K03550}.}

\bibliography{dif}

\end{document}


\title{
{Scale Separation of Shear-induced Criticality in Glasses} --- supplemental material
}

\author{Norihiro Oyama}
\email{Norihiro.Oyama.vb@mosk.tytlabs.co.jp}
\affiliation{Toyota Central R\&D Labs., Inc., Nagakute 480-1192, Japan}

\author{Takeshi Kawasaki}
\affiliation{Department of Physics, Nagoya University, Nagoya 464-8602, Japan}

\author{Kang Kim}
\affiliation{Division of Chemical Engineering, Graduate School of Engineering Science, Osaka University, Osaka 560-8531, Japan}

\author{Hideyuki Mizuno}
\affiliation{Graduate School of Arts and Sciences, The University of Tokyo, Tokyo 153-8902, Japan}


\maketitle

\section{Parameter dependence of the average number of shear transformations}\label{sec:scaling_nst}
\begin{figure*}
\includegraphics[width=0.5\linewidth]{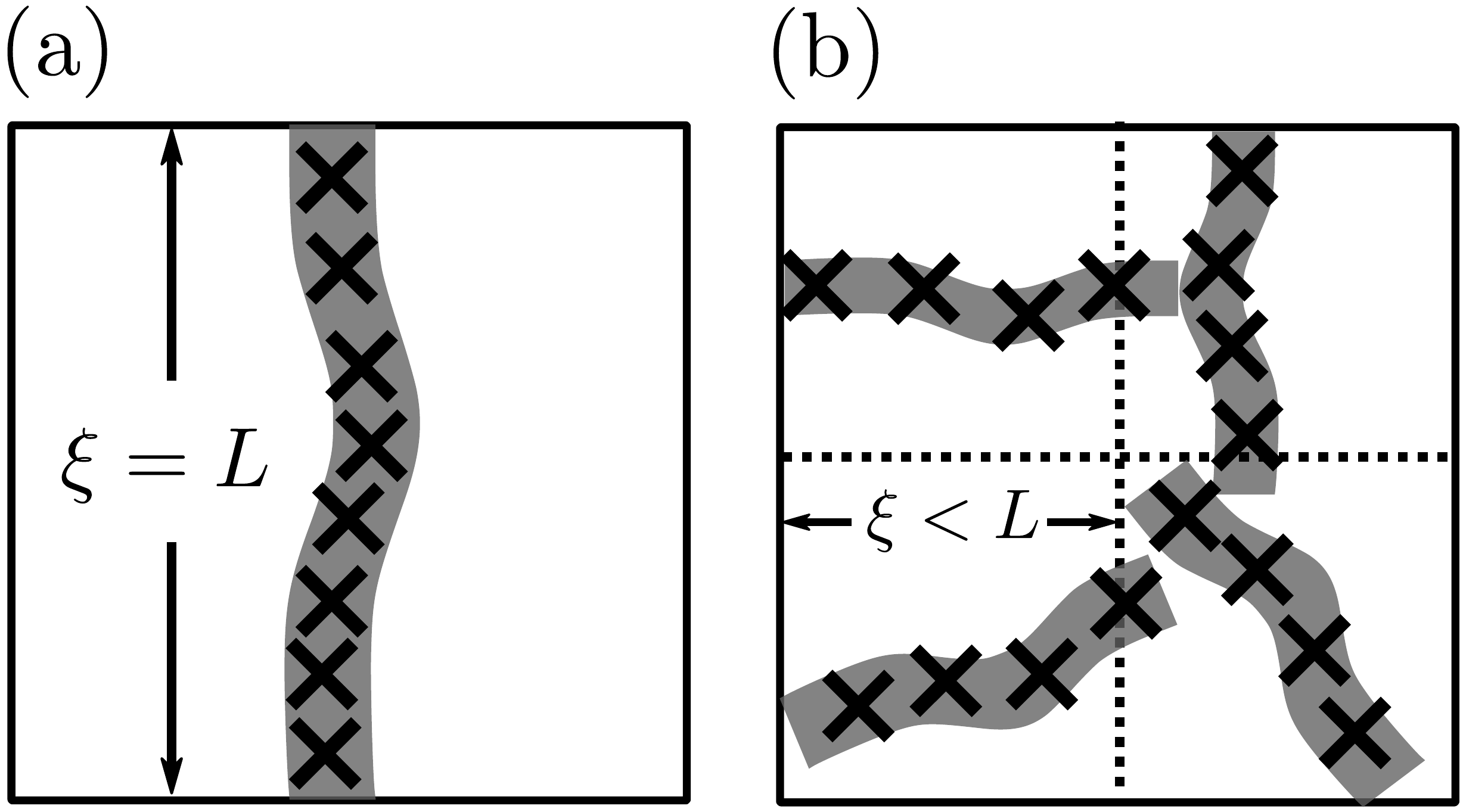}
\caption{
Sketches of quasilinear avalanches (characterized by the fractal dimension of $d_f\approx 1.0$) observed in sheared two-dimensional glasses.
(a) A system-spanning avalanche.
(b) Avalanches whose lengths $\xi$ are less than the system size $L$.
  \label{fig:sketch}}
\end{figure*}

Here, we briefly summarize the scaling argument about the parameter (shear rate $\dot{\gamma}$ and system size $N$) dependence of the average number of shear transformations (STs), denoted as $\langle N_{\rm ST}\rangle$, which was originally discussed in ref.~\cite{Oyama}.
STs are local rearrangements of particle configurations and are the elementary processes of plastic deformations.
In the sheared steady state, these STs tend to form avalanches whose spatial expansion and number depend on the shear rate $\dot{\gamma}$ in a complex manner.
In Fig.~\ref{fig:sketch}(a), we show a schematic picture of a system-spanning avalanche (where $\xi=L$).
Figure~\ref{fig:sketch}(b) depicts, on the other hand, a sketch of avalanches with smaller correlation lengths than the system linear dimension: $\xi<L$~\cite{Lin2014,Oyama}.
As expressed in Fig.~\ref{fig:sketch}b, the correlation length $\xi$ becomes smaller than $L$ when multiple avalanches are present simultaneously.

The total number of STs can be expressed as the product of the number of avalanches ($N_{\rm ava}$) and the number of STs in each avalanche ($N_{\rm ST/ava}$) as $\langle N_{\rm ST}\rangle=N_{\rm ava}\cdot N_{\rm ST/ava}$.
From the definition shown in Fig.~\ref{fig:sketch}b, $N_{\rm ava}$ can be expressed as $N_{\rm ava}=(L/\xi)^d$~\cite{Lin2014,Oyama}. 
Assuming that STs are distributed with similar spatial intervals when they form avalanches (this must statistically be the case~\cite{Oyama,LC2009}), we can estimate the number of STs in an avalanche with the size $\xi$ as $N_{\rm ST/ava}\sim\xi^{d_f}$.
Therefore, we can describe the average number of STs as $\langle N_{\rm ST}\rangle\sim (L/\xi)^d\cdot \xi^{d_f}\sim L^d\cdot\xi^{-(d-d_f)}$.
Moreover, using the following equations (scaling ansatzes and a hyperscaling relation obtained based on the statistical tilt symmetry),
\begin{align}
 \xi&\sim\Delta\sigma^{-\nu},\label{eq:xi}\\
 \dot{\gamma}&\sim\Delta\sigma^\beta,\label{eq:g_dot}\\
 \nu&=1/(d-d_f),\label{eq:nu}
\end{align}
we finally obtain this relation:
\begin{align}
    \langle N_{\rm ST}\rangle\sim N\cdot\dot{\gamma}^{1/\beta},
\end{align}
where we introduced $\Delta\sigma\equiv\sigma-\sigma_{\rm Y}$ using the yield stress $\sigma_{\rm Y}$ and used $N\sim L^d$.
Introducing the density of STs as $\langle n_{\rm ST}\rangle\equiv\langle N_{\rm ST}\rangle/N$, we obtain:
\begin{align}
    \langle n_{\rm ST}\rangle\sim \dot{\gamma}^{1/\beta}.\label{eq:ndagger}
\end{align}
This relation indicates that $\langle n_{\rm ST}\rangle$ does not show any system size dependence.

Importantly, as shown in Fig.~2(a,b) in the main text, the average number of STs $\langle N_{\rm ST}\rangle$ behaves in the same power-law manner for the entire parameter space even when $\langle N_{\rm ST}\rangle<1$.
This means that a small system that rarely has active STs behaves like a subsystem of a larger system where several STs are active at any moment.

\section{Finite Size Scaling}
\subsection{Stress response}\label{sec:stress}
To conduct finite size scaling for the stress response, by introducing a scaling function $f_\sigma(x)$, we rewrite Eq.~\ref{eq:g_dot} as:
\begin{align}
    \Delta\sigma&=\dot{\gamma}^{1/\beta}f_\sigma(\dot{\gamma}/\dot{\gamma}_{C_\sigma}),\label{eq:g_dot_2}\\
    \Leftrightarrow \Delta\sigma/\dot{\gamma}_{C_\sigma}^{1/\beta}&=(\dot{\gamma}/\dot{\gamma}_{C_\sigma})^{1/\beta}f_\sigma(\dot{\gamma}/\dot{\gamma}_{C_\sigma}),\label{eq:stress_scaling}
\end{align}
where, $\dot{\gamma}_{C_\sigma}$ stands for the characteristic shear rate at which the correlation length $\xi$ becomes exactly the same as the system linear dimension $L$ (of a given system with size $N$; thus, $\dot{\gamma}_{C_\sigma}$ is a function of $N$).
From Eqs.~\ref{eq:xi} and \ref{eq:g_dot}, we obtain $\dot{\gamma}_{C_\sigma}\sim L^{-\beta/\nu}$.
Considering this relation and Eq.~\ref{eq:stress_scaling}, we obtain:
\begin{align}
    \Delta\sigma\cdot L^{1/\nu}\sim(\dot{\gamma}\cdot L^{\beta/\nu})^{1/\beta}f_\sigma(\dot{\gamma}/L^{\beta/\nu}).
\end{align}
This means that the stress responses for different system sizes can be collapsed onto a single master curve if we plot $\Delta\sigma\cdot L^{1/\nu}$ as a function of $\dot{\gamma}\cdot L^{\beta/\nu}$.
In Fig.~2(c,d) in the main text, we demonstrated that this scaling led to a perfect collapse with the values of $\sigma_{\rm Y}$, $\nu$, and $\beta$ estimated by quasistatic simulations~\cite{Oyama2021PRE} and the slope of $\langle n_{\rm ST}\rangle$.

\subsection{Diffusion constant}\label{sec:scaling_D}
We introduced, in the main text, the scaling ansatz for the diffusion constant as:
\begin{align}
    \hat{D}\sim\Delta\sigma^{-\alpha}.\label{eq:D}
\end{align}
With Eq.~\ref{eq:g_dot}, this leads to
\begin{align}
    \hat{D}\sim\dot{\gamma}^{-\alpha/\beta}.\label{eq:D_g}
\end{align}
Using this relation, we rewrite the expression of $\hat{D}$ by introducing a scaling function $f_{\hat{D}}(x)$ as:
\begin{align}
    \hat{D}&=\dot{\gamma}^{-\alpha/\beta}f_{\hat{D}}(\dot{\gamma}/\dot{\gamma}_{C_{\hat{D}}}),\\
    \Leftrightarrow\hat{D}/\dot{\gamma}_{C_{\hat{D}}}^{-\alpha/\beta}&=(\dot{\gamma}/\dot{\gamma}_{C_{\hat{D}}})^{-\alpha/\beta}f_{\hat{D}}(\dot{\gamma}/\dot{\gamma}_{C_{\hat{D}}}),
\end{align}
where, $\dot{\gamma}_{C_{\hat{D}}}$ is the shear rate at which the length scale ${\cal L}$ governing the criticality of $\hat{D}$ spans the whole system.
We introduce a new critical exponent $\nu^\ast$ as ${\cal L}\sim\Delta\sigma^{-\nu^\ast}$.
As we discuss in the main text, ${\cal L}$ can be both $\xi$ and $\zeta$ (thus, correspondingly $\nu^\ast$ can be both $\nu$ and $\tilde{\nu}$) depending on the microscopic dissipation mechanism.
Using this exponent $\nu^\ast$, $\dot{\gamma}_{C_{\hat{D}}}$ is expressed as $\dot{\gamma}_{C_{\hat{D}}}\sim L^{-\beta/\nu_\ast}$, and the following relation can be derived:
\begin{align}
    \hat{D}/ L^{\alpha/\nu^\ast}=(\dot{\gamma}\cdot L^{\beta/\nu^\ast})^{-\alpha/\beta}f_{\hat{D}}(\dot{\gamma}/\dot{\gamma}_{C_{\hat{D}}}).
\end{align}
From Eq.~\ref{eq:D} and the scaling ansatz ${\cal L}\sim\Delta\sigma^{-\nu^\ast}$, we obtain:
\begin{align}
    \hat{D}\sim {\cal L}^{\alpha/\nu^\ast}.
\end{align}
This means that $L^{\alpha/\nu^\ast}$ should obey the same scaling as $\hat{D}_0$ (in accordance with the main text, we refer to the plateau value of $D$ below $\dot{\gamma}_{C_{\hat{D}}}$ as $\hat{D}_0$).
Since $\hat{D}_0$ scales linearly with $L$, we obtain a relation $L^{\alpha/\nu^\ast}\sim\hat{D}_0\sim L$ (remember the relation $\hat{D}_0\sim L$).
In other words, we obtain a scaling relation
\begin{align}
    \alpha=\nu^\ast.\label{eq:alpha_nu}
\end{align}

{From the above discussion, we obtain:
\begin{align}
    \hat{D}/ L&=(\dot{\gamma}\cdot L^{\beta/\nu^\ast})^{-\nu^\ast/\beta}f_{\hat{D}}(\dot{\gamma}/\dot{\gamma}_{C_{\hat{D}}})\\
    &=(L\cdot\dot{\gamma}^{\nu^\ast/\beta})^{-1}f_{\hat{D}}(\dot{\gamma}/\dot{\gamma}_{C_{\hat{D}}}).
\end{align}
This relation indicates that the strain-based diffusion coefficient $\hat{D}$ can be collapsed by scaling $\hat{D}/L$ and $L\cdot\dot{\gamma}^{{\nu^\ast}/\beta}$ using the critical exponent $\nu^\ast$ corresponding to the true critical correlation length ${\cal L}$ governing $\hat{D}$.
We note that this scaling form is the same as the one introduced in ref.~\cite{LC2009} (in this reference, this relation was derived from a phenomenological argument as shown in Sec.~\ref{sec:derivation}) and we used this form in Figs.~3 and 5 in the main text.
}

\section{Damping mechanism dependence of yielding criticality}\label{sec:damping}
In refs.~\cite{LC2009,Roy}, the shear rate $\dot{\gamma}$ and system size $N$ dependence of both the stress response $\langle\sigma\rangle$ and the strain-based diffusion coefficient $\hat{D}$ have been measured for two-dimensional glass systems.
In ref.~\cite{LC2009}, an underdamped system was considered where the dissipation was introduced in the form of interparticle dissipative forces (the same as the one in our System A).
In ref.~\cite{Roy}, on the other hand, an overdamped system is treated.
Both of these studies have reported that the exponents $1/\beta$ and $\alpha/\beta$ appearing in Eqs.~\ref{eq:g_dot_2} and \ref{eq:D_g} are identical (the values of exponents are different between these two systems: they are 1/2 in ref.~\cite{LC2009}, while they are 1/3 in ref.~\cite{Roy}).
This means that a relation $\alpha\approx 1.0$ holds, and moreover, because of a relation $\alpha=\nu^\ast$ (Eq.~\ref{eq:alpha_nu}), the critical exponent for the correlation length that governs the criticality of $\hat{D}$ should be ${\nu}^\ast\approx 1.0$ in these systems~\cite{LC2009,Roy}.

Regarding the avalanche fractal dimension $d_f$, the precise values were not reported in refs.~\cite{LC2009,Roy}.
However, for various sheared two-dimensional glass systems, values of approximately $d_f\approx 1.0$ have been reported universally~\cite{ML2004b,Zhang2017,Oyama2019PRL,Oyama2021PRE}.
If we assume $d_f\approx 1.0$ as well for the systems in refs.~\cite{LC2009,Roy}, this value leads to $\nu\approx 1.0$ from the scaling relation $\nu=1/(d-d_f)$.
In conjunction with the discussion in the previous paragraph, this means $\nu^\ast\approx \nu\approx 1.0$ and there is no scale separation between the criticality of the stress and the diffusion constant in these systems. 

As discussed in the main text, the scale separation is absent in System A (and thus, in the system of ref.~\cite{LC2009} as well) because the high-wavenumber local dynamics are damped.
Similarly, the high-wavenumber local dynamics are suppressed in the overdamped system in ref.~\cite{Roy} (in the overdamped situation, dynamics at all length scales are totally damped) and the scale separation is absent as well.

\section{Scaling relation for critical exponents of length scales}
In this section, we explain how we can extract the critical exponents that characterize two critical correlation lengths $\xi$ (that of avalanches) and $\zeta$ (that of STs) from the average number of STs $\langle N_{\rm ST}\rangle$.

\subsection{Correlation length of avalanches}
The avalanche correlation length corresponds to, by definition, the linear dimension of a system in which only one single avalanche fits ($N_{\rm ava}=1$).
Therefore, if we consider a situation where $L=\xi$, $\langle N_{\rm ST}\rangle=N_{\rm ST/ava}\sim\xi^{d_f}$ is satisfied (see Fig.~\ref{fig:sketch}b and the discussion in Sec.~\ref{sec:stress}).
Then, from $\langle N_{\rm ST}\rangle\sim N\cdot \dot{\gamma}^{1/\beta}\sim L^d\cdot\Delta\sigma$, we obtain a relation that is totally consistent with Eq.~\ref{eq:nu} as (using the conditions $L=\xi$, $\langle N_{\rm ST}\rangle=\xi^{d_f}$):
\begin{align}
    &\ \ \langle N_{\rm ST}\rangle\sim L^d\Delta\sigma\\
    &\Leftrightarrow\xi^{d_f}\sim\xi^d\cdot\Delta\sigma\\
    &\Leftrightarrow\xi\sim\Delta\sigma^{-1/(d-d_f)}\\
    &\therefore \nu=1/(d-d_f)
\end{align}

\subsection{Correlation length of shear transformations}
Here, we consider the second correlation length $\zeta$.
This length is defined as the linear dimension of purely elastic subdomains compartmentalized by STs.
Such a length scale corresponds to the linear dimension of the average volume where only one ST is enclosed\footnote{The correspondence must be easily understood by considering a 1D system. Under periodic boundary conditions, the average spacing between STs and the average spanning occupied by each ST are the same.}.
Therefore, we call $\zeta$ the ST correlation length.
Similar to the one for $\xi$, we set a scaling ansatz for $\zeta$ as $\zeta\sim\Delta\sigma^{-\tilde{\nu}}$, introducing another critical exponent $\tilde{\nu}$.
With this definition of $\zeta$, if we consider a situation with $L=\zeta$, we expect $\langle N_{\rm ST}\rangle=1$.
Then, $\langle N_{\rm ST}\rangle\sim L^d\cdot\Delta\sigma$ can be recast into (using $L=\zeta$):
\begin{align}
    \langle N_{\rm ST}\rangle&\sim L^d\Delta\sigma\\
    \Leftrightarrow1&\sim \zeta^d \cdot\Delta\sigma\\
    \Leftrightarrow\zeta&\sim\Delta\sigma^{-1/d}
\end{align}
and we obtain the following relation:
\begin{align}
    \tilde{\nu}=1/d
\end{align}

\section{Phenomenological description of the  criticality of the diffusion constant}\label{sec:derivation}
In Sec.~\ref{sec:original}, we recapitulate the phenomenological derivation of the criticality of $\hat{D}$ originally proposed in ref.~\cite{LC2009}.
In Sec.~\ref{sec:modification}, we also show that, by making minor corrections to the derivation of the quantitative prediction for the critical exponent $n_{\hat{D}}$, we can derive an identical hyperscaling relation that is obtained by the framework of the yielding criticality.

In Secs.~\ref{sec:original} and \ref{sec:modification}, the screening effect is assumed to be negligible.
In Sec.~\ref{sec:when_screened}, we discuss the modifications needed for the systems where screening effects play a major role.

\subsection{Eshelby fields as the origin of diffusive motions}\label{sec:original}
As in the main text, we consider a two-dimensional system with shear applied in the $x$ direction and we derive a theoretical expression for the strain-based diffusion constant in the $y$-direction.
Since the displacement in the $y$-direction is mainly induced by avalanches (of STs), we first introduce the $y$-component of the displacement of particle $i$ due to the $\alpha$th avalanche as $u_{y,i}(\alpha)$, which is described as:
\begin{align}
u_{y,i}(\alpha)=\sum_{k\in{\alpha}} u_{y,k}^{\rm E}(\boldsymbol{r}_i-\boldsymbol{r}_k^{\rm ST}),
\end{align}
where $\boldsymbol{r}_i$ and $\boldsymbol{r}_{k}^{\rm ST}$ represent the positions of particle $i$ and $k$th ST belonging to the $\alpha$th avalanche respectively, and $u_{y,k}^{\rm E}$ is the Eshelby displacement field that the $k$th ST generates.
We will present the precise form of $u_{y,k}^{\rm E}$ later (in Eq.~\ref{eq:fourier_uye}).
The summation $\sum_{k\in\alpha}$ runs over all STs forming the $\alpha$th avalanche.
Now we introduce the density field of the ST centers as
\begin{align}
\phi_\alpha(\boldsymbol{r})=\sum_{k\in{\alpha}}\delta(\boldsymbol{r}-\boldsymbol{r}_k^{\rm ST}).
\end{align}
Then, using $\phi_\alpha$, we can write down the displacement and its square as:
\begin{align}
u_{y,i}(\alpha)&=\int d\boldsymbol{r}_{\rm ST}[\phi_\alpha(\boldsymbol{r}_{\rm ST})\times u_{y,k}^{\rm E}(\boldsymbol{r}_i-\boldsymbol{r}_{\rm ST})],\\
u^2_{y,i}(\alpha)&=\int d\boldsymbol{r}_{\rm ST}d\boldsymbol{r}_{\rm ST^\prime}
[\phi_\alpha(\boldsymbol{r}_{\rm ST})\phi_\alpha(\boldsymbol{r}_{\rm ST^\prime})\notag\\
&\ \ \ \ \ \times u_{y,k}^{\rm E}(\boldsymbol{r}_i-\boldsymbol{r}_{\rm ST})
u_{y,k}^{\rm E}(\boldsymbol{r}_i-\boldsymbol{r}_{\rm ST^\prime})]\notag\\
&=\int d\boldsymbol{r}d\boldsymbol{R}
[\phi_\alpha(\boldsymbol{r})\phi_\alpha(\boldsymbol{r}-\boldsymbol{R})\times u_{y,k}^{\rm E}(\boldsymbol{r})
u_{y,k}^{\rm E}(\boldsymbol{r}-\boldsymbol{R})]\label{eq:squared_uyi}.
\end{align}
We used the translational invariance for the last equality.
Using these relations, we can write the particle-averaged squared displacement in the $y$-direction as:
\begin{align}
    \overline{u_y^2(\alpha)}=\frac{1}{L^2}\int d\boldsymbol{r}_i u_{y,i}^2(\alpha).\label{eq:global_usq}
\end{align}

Now we introduce the displacement correlation function $\Gamma(\boldsymbol{R})$ as:
\begin{align}
    \Gamma(\boldsymbol{R})
    &=\int d\boldsymbol{r} u_y^{\rm E}(\boldsymbol{r})u_y^{\rm E}(\boldsymbol{r}-\boldsymbol{R})\\
    &=\frac{1}{(2\pi)^2}\int d\boldsymbol{q}|\hat{u}_y^{\rm E}(\boldsymbol{q})|^2\exp[{-i\boldsymbol{q}\cdot\boldsymbol{R}}].
\end{align}
In the Fourier space, the Eshelby displacement field can be expressed as:\footnote{The Fourier transform of the Eshelby field $\hat{u}_y^{\rm E}(\boldsymbol{q})$ is $\hat{u}_y^{\rm E}(\boldsymbol{q})=-2\mu(\hat{O}_{xy}\cdot iq_y+\hat{O}_yy\cdot iq_x)\hat{\epsilon}_{xy}^{\rm pl}$. Here, we used the Oseen tensor $\hat{\boldsymbol{O}}(\boldsymbol{q})=\frac{1}{\mu q^2}\left(\boldsymbol{I}-\frac{\boldsymbol{qq}}{q^2}\right)$~\cite{Picard2004} with the shear modulus $\mu$ and a plastic strain source $\epsilon_{xy}^{\rm pl}=\Delta\epsilon_0 a^d\delta(\boldsymbol{r}_{\rm ST})$ ($d=2$ is the spatial dimension).}
\begin{align}
    \hat{u}_y^{\rm E}(\boldsymbol{q})=-ia^2\Delta\epsilon_0\frac{q_x(q_x^2-q_y^2)}{q^4},\label{eq:fourier_uye}
\end{align}
and we can write the correlation function as:
\begin{align}
    \Gamma(\boldsymbol{R})=\frac{a^4\Delta\epsilon_0^2}{4\pi^2}
    \int_{q_{\rm min}}^\infty\frac{dq}{q}\int_0^{2\pi}d\theta^\prime\gamma(\theta^\prime)e^{-iqR\cos(\theta^\prime-\theta)},
\end{align}
where we introduced $\boldsymbol{R}=(R,\theta)$, $q_{\rm min}\sim 1/L$, and $\gamma(\theta^\prime)=\cos^2\theta^\prime(\cos^2\theta^\prime-\sin^2\theta^\prime)^2$.
Thus, integrating over $\theta$ and introducing $G(z,\theta)=2J_0(z)-3\cos(2\theta)J_2(z)+2\cos(4\theta)J_4(z)-\cos(6\theta)J_6(z)$ with $J_n$ the Bessel functions, we obtain
\begin{align}
    \Gamma(\boldsymbol{R})=\frac{a^4\Delta\epsilon_0^2}{16\pi}\int_{R/L}^\infty \frac{dz}{z}G(z,\theta).\label{eq:Gamma_final}
\end{align}

Because, as discussed in Sec.~\ref{sec:damping}, the fractal dimension of avalanches is universally $d_f\approx 1.0$ in 2D systems, avalanches have quasilinear structures.
We assume that, along these quasilinear avalanches, shear transformations are distributed uniformly with a linear density $\psi$.
Then, using $\Gamma(\boldsymbol{R})$ and Eq.~\ref{eq:global_usq}, we can write the square of the sum of the Eshelby fields generated by the $\alpha$th avalanche as:
\begin{align}
    \overline{(u_y^{\alpha})^2}=\frac{\psi^2}{L^2}\int_0^\xi\int_0^\xi dxdx^\prime\Gamma[(x-x^\prime)\boldsymbol{e}_x].\label{eq:usq_a,x}
\end{align}
Here, $\boldsymbol{e}_x$ is the $x$-directional unit vector.
For simplicity, we assumed the avalanche of the correlation length\footnote{Although it is quite nontrivial whether the correlation length here is identical to the avalanche correlation length $\xi$, it is indeed identical, as discussed in Sec.~\ref{sec:modification}. For simplicity, therefore, we use $\xi$ here.} $\xi$ is located along the $x$-direction and is near the origin ($0\le x\le \xi, y=0$).
This simplification does not detract from the generality but locates the integration in Eq.~\ref{eq:usq_a,x} solely along the avalanche.
Finally, inserting Eq.~\ref{eq:Gamma_final} into \ref{eq:usq_a,x}, we obtain
\begin{align}
\overline{(u_y^{\alpha})^2}\approx \frac{a^4\Delta\epsilon^2_0\psi^2}{2\pi} \frac{\xi^2}{L^2}\ln(\frac{L}{\xi}),\label{eq:u_sq}
\end{align}
as the leading-order expansion in $\xi/L$.

Using the number of avalanches $N_A$ induced during a strain interval of $\delta\gamma$, the (strain-based) diffusion constant $\hat{D}$ can be expressed as
\begin{align}
    \hat{D}=\frac{N_A \overline{(u_y^{\alpha})^2}}{2\delta\gamma}.
\end{align}
Since the number of STs per avalanche~\footnote{Strictly, the number of STs per avalanche $N_{\rm ST/ava}$ is proportional to $\xi^{d_f}$, as discussed in Sec.~\ref{sec:scaling_nst}. If we employ this strict scaling, we obtain $\hat{D}_0\sim L^{d-d_f}$, which is derived in ref.~\cite{Karimi}. However, we used an approximated description here, obeying the original discussion in ref.~\cite{LC2009}.} is proportional to $\psi \xi$, $N_A$ per $\delta\gamma$ can be described by using the number of ST $N_f$ during $\delta\gamma$ as:
\begin{align}
    N_A = \frac{N_f}{\psi \xi}.
\end{align}
Moreover, the average plastic strain emitted by STs is described as $a^2\Delta\epsilon_0/L^2$, and we can derive a relation for $N_f$ based on the balance between the total plastic strain induced by STs and the externally applied strain as:
\begin{align}
    N_f = \frac{L^2\delta\gamma}{a^2\Delta\epsilon_0}.\label{eq:N_f}
\end{align}
From Eqs.~\ref{eq:u_sq}-\ref{eq:N_f}, we finally express $\hat{D}$ as:
\begin{align}
    \hat{D}\approx\frac{a^2\Delta\epsilon_0}{4\pi}\psi\xi\ln\left(L/\xi\right).
\end{align}
This explains the linear dependence of $\hat{D}$ on the correlation length $\xi$ (or, the finite size effect).

\subsection{Relation to yielding criticality}\label{sec:modification}
In this subsection, we first explain how a relation $\xi\sim \dot{\gamma}^{1/2}$ was phenomenologically derived\footnote{Again, we stress that in the original discussion in ref.~\cite{LC2009}, the correlation length was not tightly associated with avalanche correlation length.} in ref.~\cite{LC2009}.
Then, we explain that, with minor corrections for the treatment of the correlation length, this prediction based on a phenomenological theory becomes fully identical to the hyperscaling relation derived from scaling descriptions based on yielding criticality.
Therefore, the updated phenomenological theory provides the correct estimation for the diverse relation between $\xi$ and $\dot{\gamma}$, although its applicability is limited to the situation where the scale separation is absent (or, the screening effects are negligible).

In ref.~\cite{LC2009}, the relation $\xi\sim \dot{\gamma}^{1/2}$ was derived in a way that is independent of the system details: the authors compared the average occurrence rate of STs and their lifetime.
Since, as introduced above, the average plastic strain induced by a single ST is written as $a^2\Delta\epsilon_0/L^2$, the average occurrence rate ${\cal R}$ of STs under a given shear rate $\dot{\gamma}$ is expressed as (obtained from a simple balance equation between externally applied strain per time and realized plastic strain):
\begin{align}
    {\cal R}=\frac{L^2\dot{\gamma}}{a^2}\Delta\epsilon_0.
\end{align}
Then, obeying the definition of $\xi$ shown in Fig.~\ref{fig:sketch}, the occurrence rate of STs belonging to a single avalanche 
${\cal R}_{\xi}$ can be expressed as:
\begin{align}
    {\cal R}_\xi=\frac{\xi^2}{L^2}{\cal R}=\frac{\xi^2\dot{\gamma}}{a^2}\Delta\epsilon_0.
\end{align}
The authors of ref.~\cite{LC2009} further considered that ``near-field signals must not overlap''.
In other words, they assumed that the occurrence rate should be nearly equal\footnote{In the original discussion in ref.~\cite{LC2009}, this equality was treated as an inequality as ${\cal R}_\xi\le \tau^{-1}$.} to the inverse lifetime of STs $\tau^{-1}$:
\begin{align}
{\cal R}_\xi\approx \tau^{-1}.
\end{align}
Finally, we obtain:
\begin{align}
    \xi\approx \sqrt{\frac{a^2\Delta\epsilon_0}{\dot{\gamma}\tau}}\sim \dot{\gamma}^{-1/2}.
\end{align}
Since the derivation did not rely on any system-dependent details, this square-root power-law relation has been believed to be universally applicable to any system.
However, the measurement result in ref.~\cite{Roy} was largely different: $\xi\sim\dot{\gamma}^{-1/3}$.

In fact, to derive a correct prediction, we need to consider the balance between the lifetime and occurrence rate of avalanches, not STs.
Since the number of STs\footnote{Strictly, $\psi$ here should be different from the one introduced in Eq.~\ref{eq:usq_a,x} because, in Eq.~\ref{eq:usq_a,x}, $d_f=1.0$ was inserted. We employed the same notation for simplicity, because $d_f\approx 1.0$.} composing an avalanche of length $\xi$ is $N_{\rm ST/ava}={\psi}\xi^{d_f}$, the total plastic strain $\gamma_{\xi}$ provided by an avalanche is:
\begin{align}
    \gamma_\xi=N_{\rm ST/ava}\cdot\frac{a^2}{L^2}\Delta\epsilon_0=\psi\xi^{d_f}\frac{a^2}{L^2}\Delta\epsilon_0.
\end{align}
Therefore, the occurrence rate of STs belonging to an avalanche should be expressed as:
\begin{align}
    {\cal R}_\xi=\frac{\xi^2\dot{\gamma}}{a^2\psi\xi^{d_f}}\Delta\epsilon_0.\label{eq:update_R}
\end{align}
On the other hand, the avalanche of length $\xi$ has a size-dependent lifetime (this is a common scaling ansatz employed for yielding criticality~\cite{Lin2014}):
\begin{align}
    \tau_\xi=\xi^z.\label{eq:update_tau}
\end{align}
Equating Eqs~\ref{eq:update_R} and \ref{eq:update_tau}, we obtain:
\begin{align}
z=(\beta-1)/\nu.\label{eq:hyperscaling_z}
\end{align}
This is exactly the hyperscaling relation that we can derive based on the statistical tilt symmetry of the governing equation~\cite{Lin2014}.
At the same time, from Eq.~\ref{eq:update_R} and \ref{eq:update_tau}, we also obtain:
\begin{align}
\xi\sim\dot{\gamma}^{-1/(z+d-d_f)}.\label{eq:xi_gdot}
\end{align}
Eqs.~\ref{eq:hyperscaling_z} and \ref{eq:xi_gdot} provide predictions that are totally consistent with numerical observations for systems in both refs.~\cite{LC2009,Roy} as explained below.
\begin{enumerate}
    \item In the case of ref.~\cite{LC2009}\\
    Inserting the numerical result $\beta=2$ and an assumption $\nu=1$ (see Sec.~\ref{sec:damping}) into Eq.~\ref{eq:hyperscaling_z}, we obtain $z=1$.
    From this value and Eq.~\ref{eq:xi_gdot}, we obtain a relation $\xi\sim\dot{\gamma}^{-1/2}$, which is in line with the numerical result.
    \item In the case of ref.~\cite{Roy}\\
    Similarly, inserting the observation results $\beta=3$ and $\nu=1$ into Eq.~\ref{eq:hyperscaling_z}, we obtain $z=2$, a largely different value from that for ref.~\cite{LC2009}.
    From this value and Eq.~\ref{eq:xi_gdot}, we obtain $\xi\sim\dot{\gamma}^{-1/3}$.
    This is, again, in accordance with the numerical result in ref.~\cite{Roy}.
\end{enumerate}

\subsection{When screening effects cannot be ignored}\label{sec:when_screened}
In Secs.~\ref{sec:original} and \ref{sec:modification}, we assumed that the screening effect is negligible.
When the effects of the screening cannot be ignored, we need to employ the ST correlation length $\zeta$ in Eq.~\ref{eq:usq_a,x} instead of the avalanche correlation length $\xi$.
Then, we simplyobtain a relation:
\begin{align}
    \hat{D}\approx\frac{a^2\Delta\epsilon_0}{4\pi}\psi\zeta\ln\left(L/\zeta\right).
\end{align}

Regarding the contents of Sec.~\ref{sec:modification}, Eqs.~\ref{eq:hyperscaling_z} and \ref{eq:xi_gdot}, the mains results, are both applicable even when the screening effects play a major role.
However, we cannot extract the scaling relation for $\zeta$ from these equations, and thus we need to rely on other approaches.
We proposed one exampe of such approaches in the main text.

\section{Summary of Critical Exponents}
We summarized the critical exponents in the systems of our work and refs.~\cite{LC2009,Roy} in Table~\ref{table:exponents}.
\tabcolsep = 16pt
\begingroup
\renewcommand{\arraystretch}{1.5}
\begin{table}[tb]
  \caption{{Values of the critical exponents for different systems}}
  \centering\label{table:exponents}
  \begin{tabular}{c||c|c|c|c|c|c}
    \hline
     & $\beta$ & $d_f$ & $\nu$ & $\tilde{\nu}$  & $\alpha$ & $n_{\hat{D}}$\\
    \hline \hline
    Our work (System A) & 1.39 &\multirow{2}{*}{$\approx 1.03$} & \multirow{2}{*}{$\approx 1.04$} & \multirow{4}{*}{0.5}  &$\approx 1.0$ &0.72\\\cline{1-2}\cline{6-7}
    Our work (System B) &1.32 &  &  &  &0.5 &0.38\\\cline{1-4}\cline{6-7}
    Ref.~\cite{LC2009} &2.00  &\multirow{2}{*}{$\approx 1.0$*}  &\multirow{2}{*}{$\approx 1.0$*} & &\multirow{2}{*}{$\approx 1.0$*} &0.5\\\cline{1-2}\cline{7-7}
    Ref.~\cite{Roy} &3.00  &  & & & &0.33\\
    \hline
    \multicolumn{7}{r}{*these values are estimated according to the discussion in Sec.~\ref{sec:damping}}\\
  \end{tabular}
\end{table}
\endgroup

\section{Mobile particles}
In Fig.~4 in the main text, ``mobile particles'' are highlighted in red to visualize the spatial structure of STs (that are identified by Im-INMs) in an intuitively easy-to-understand manner.
Mobile particles are selected according to the participation ratio explained below.
The participation ratio $p_k$ of the $k$-th INM is defined as:
\begin{align}
    p_k = \frac{(\Sigma_i|\boldsymbol{e}_{k,i}|^2)^2}{N\Sigma_i|\boldsymbol{e}_{k,i}|^4},    
\end{align}
where $\boldsymbol{e}_{k,i}$ is the component corresponding to particle $i$ in the eigenvector of the $k$-th INM $\boldsymbol{e}_k\equiv \{ \boldsymbol{e}_{k,1}, \boldsymbol{e}_{k,2}, ... , \boldsymbol{e}_{k,N} \}$.
This participation ratio $p_k$ provides a fraction of mobile particles in the $k$-th INM without introducing any extra parameters:
The value of $p_k=1$ indicates that all particles have exactly the same magnitude of $\boldsymbol{e}_{k,i}$, while $p_k=1/N$ means that only one particle has nonzero components.

We selected the particles that have the $p_k\cdot N$ largest magnitudes of the particle-based eigenvector $|\boldsymbol{e}_{k,i}|$ as mobile particles of the $k$-th INM.

\bibliography{dif}